\documentstyle[twocolumn,prd,aps]{revtex}  
\input epsf
\begin{document}                
\title{Tracking Black Holes in Numerical Relativity}
\author{Scott A. Caveny, Matthew Anderson, Richard A. Matzner}
\address{Center for Relativity, The University of Texas at Austin, Austin, 
TX 78712-1081 }
%
\maketitle
\begin{abstract}                
This work addresses and solves the problem of generically tracking
black hole event horizons in computational simulation of black hole
interactions. Solutions
of the hyperbolic eikonal equation, solved on a curved spacetime
manifold containing black hole sources, are employed in
development of a robust tracking method capable of continuously
monitoring arbitrary changes of topology in the event horizon, as
well as arbitrary numbers of gravitational sources. The method
makes use of continuous families of level set viscosity solutions
of the eikonal equation with identification of the black hole
event horizon obtained by the signature feature of discontinuity
formation in the eikonal's solution. The method is employed in the
analysis of the event horizon for the asymmetric merger in a
binary black hole system. In this first such three dimensional analysis, we
establish both
qualitative and quantitative physics for the asymmetric collision;
including: 1. Bounds on the topology of the throat connecting the
holes following merger, 2. Time of merger, and 3. Continuous
accounting for the surface of section areas of the black hole
sources.
\end{abstract}
\section{Introduction}               
The idea of a totally collapsed gravitational source, from which
nothing --- not even light --- can escape, is an old one, dating
back at least to the work of Laplace (and others) in the
eighteenth century \cite{he:}. Since those original
considerations, research has uncovered a wealth of understanding
regarding the physics of black holes. In general, the work has
followed two major routes,  with an ever - narrowing gap between
the approaches. Along one direction, black holes are studied as
mathematical solutions in a given theory of gravity including
Newtonian theory, post-Newtonian metric theories of gravity,
Einsteinian gravity, as well as semi-classical and (more recently)
various attempts at full quantum theories of gravity. The second
direction considers black holes in the astrophysical and
astronomical contexts. Both fields of research have seen
increasing activity over the years and have made startling discoveries
concerning the physics of black holes; including the gravitational
collapse theory of Oppenheimer  and Snyder \cite{opsny:}, proofs
of uniqueness and stability of black holes in Einstein's general
relativity \cite{carter:}, thermodynamic properties of black holes
including the three laws of black hole mechanics \cite{bch:} and
mechanisms for black hole radiation and evaporation \cite{wald2:},
discovery of critical phenomena in black hole formation
\cite{choptuik:}, experimental identification of both
astrophysical black hole sources themselves \cite{as1:} and their
event horizons \cite{e1:}, experimental signatures for super -
massive black holes in galactic centers \cite{ambh:}, and
experimental bounds on the distribution and spectrum of black hole
sources and collisions \cite{eb:}.

Numerical relativity, which addresses computational solution of
Einstein's equation \cite{lehner:}, is an active participant in
both the mathematical and the astrophysical programmes of research. 
With the advent of the Laser
Interferometric Gravitational Observatory and other similar efforts, 
considerable attention has focused on the
generic binary black hole coalescence (BBHC) problem, expected as
the strongest sources for gravitational waves \cite{bbhc1:}. As
discussed in \cite{lehner:}, this is a very difficult and
interesting problem that can only successfully be addressed in the
computational domain. Consequently, the binary black hole
coalescence problem has become an active subject of numerical
relativity. One particular problem in the computational domain is
the computational definition, detection and tracking of the black
hole event horizon itself \cite{eh1:}, \cite{eh2:}, \cite{eh3:},
\cite{eh4:}, \cite{eh5:}, \cite{is1:}.

The present work completely solves the problem of numerically
tracking black hole event horizons. The solution is complete in
the sense that a single method is presented such that any one
implementation of the method can generically detect arbitrary
numbers of black hole event horizons undergoing arbitrarily strong
gravitational interactions. For example, using a single computational
code of our method we analyse both single black holes and black holes
undergoing merger; and no special modifications of our code are 
required to handle these distinct dynamics. 

The present article, describing our generic method 
for tracking black hole event horizons, is divided as follows. In section II
the eikonal equation, the foundation of our method, is described in
sufficient detail to be employed in an event horizon tracker. In particular,
we focus on the signature behavior of a black hole event 
horizon in solutions of the eikonal equation. As usual in numerical work,
there are a variety of possible implementations of the approach. In section
III, several closely related systems of equations are presented and one 
particular system is singled out for consideration. The system chosen
makes use of
an explicit second order diffusion, or viscosity, term and we show in section 
III the relationship between solutions of the diffusive equations of motion
and the continuum eikonal equation of interest. Section IV details
extraction of the two dimensional sections of the event horizon for each 
time level of a numerical evolution. Such extraction is crucial for 
carrying out area, mass, and spin calculations. Section IV also shows 
the accuracy of our implementation by considering a parameter space survey 
of single Kerr black holes. Sections V - X describe in detail
the first three dimensional application of our (or any) method
to the numerical analysis of the asymmetric
binary black hole coalescence problem. In particular we search for
evidence of any nontrivial topology of the horizon immediately following 
merger.
In distinction to the prediction of \cite{wini2:} we find a topologically
spherical horizon to within the accuracy of our three dimensional mesh.
Further, area analysis of the candidate numerical event horizon is 
carried out in conjunction with analysis of the black hole 
apparent horizons, which reveals both a time of merger much earlier then 
estimated using apparent horizons and  mass energy estimates much 
larger those found using apparent horizon tracking methods. Our conclusions
are presented in section XI.

\section{The Eikonal}
To begin, since the Lagrangian $L = g_{ab}\dot{x}^{a}\dot{x}^{b}$
of null geodesic motion has only kinetic terms it is equal to the
associated Hamiltonian\footnote{We use early-latin indices to denote
spacetime components $a,b,c = 0,1,2,3$; mid-latin indices denote
spatial components $i,j,k=1,2,3$.}. Legendre transformation
\begin{equation}
H\left(\tau, x^{a},p_{b}\right) =\frac{d{x}^{c}}{d\tau} p_{c} -
L\left(\tau, x^{d},\frac{d{x}^{e}}{d\tau}\right)
\end{equation}
where
\begin{equation}
p_{a} \equiv \frac{\partial L}{\partial
\left(\frac{d{x}^{a}}{d\tau}\right)},
\end{equation}
sets $L = H$. The corresponding Hamiltonian equations are
\begin{equation}
\frac{d{x}^{a}}{d\tau} = \frac{\partial H}{\partial p_{a}} = 2
g^{ab}p_{b}
\end{equation}
\begin{equation}
\frac{d{p}_{b}}{d\tau} =  - \partial_{b} H = -
p_{c}p_{d}\partial_{b} g^{cd}.
\end{equation}

It is generic that the spacetime metric is independent of the
affine parameter $\tau$. There is thus a first integral associated
to geodesic motion. Making use of this property permits
elimination of the affine parameter in favor of coordinate time
$t$. To see this, it is convenient to adopt the ADM variables
\begin{equation}
g^{tt} = \frac{1}{\alpha^{2}},g^{ti} =
\frac{\beta^{i}}{\alpha^{2}},g^{ij} = \gamma^{ij} -
\frac{\beta^{i}\beta^{j}}{\alpha^{2}}.
\end{equation}
Here $\gamma^{ij}$ is the inverse spatial metric. 
Substituting the ADM variables directly into the Hamiltonian gives
\begin{equation}\label{6eq}
H=p_{a}g^{ab}p_{b} = -\frac{1}{\alpha^{2}}p_{t}^{2}
+\frac{2}{\alpha^{2}}\beta^{i}p_{i} + p_{i}\left(\gamma^{ij} -
\frac{1}{\alpha^{2}} \beta^{i}\beta^{j} \right)p_{j}.
\end{equation}
Since the Hamiltonian is not explicitly $\tau$ dependent  there is
a constant of the motion
\begin{equation}\label{geodesics}
-\frac{1}{\alpha^{2}}p_{t}^{2} +\frac{2}{\alpha^{2}}\beta^{i}p_{i}
+ p_{i}\left(\gamma^{ij} - \frac{1}{\alpha^{2}} \beta^{i}\beta_{j}
\right)p_{j} = \omega^{2}.
\end{equation}
For $\omega^{2} < 0 $, $\omega^{2} =0 $, or $\omega^{2} > 0 $ the
motion is said to be timelike, null, or spacelike. Without loss of
generality,  assuming null geodesic motion, solution of
(\ref{geodesics}) by ordinary algebra yields:
\begin{equation}\label{p}
p_{t} = \beta^{i} p_{i} \pm \alpha \sqrt{p_{i}\gamma^{ij}p_{j}}.
\end{equation}
With this result, the Hamiltonian can be explicitly factored:
\begin{equation}
H = H_{+}H_{-} = 0
\end{equation}
where
\begin{equation}\label{root}
H_{\pm} = p_{t} - \beta^{i}p_{i} \pm \alpha
\sqrt{p_{i}\gamma^{ij}p_{j}} = 0.
\end{equation}
In the case of either root Hamilton's canonical equations become
\begin{equation}\label{teq}
\frac{dt}{d\tau} = H_{\mp}\frac{\partial H_{\pm}}{\partial p_{t}}
= H_{\mp}
\end{equation}
\begin{equation}\label{2eq}
\frac{dx^{i}}{d\tau} = H_{\mp}\frac{\partial H_{\pm}}{\partial
p_{i}} = H_{\mp}\left(-\beta^{i} \pm \alpha
\frac{\gamma^{ij}p_{j}}{\sqrt{p_{k}\gamma^{kl}p_{l}}}\right)
\end{equation}
\begin{equation}
\frac{dp_{t}}{d\tau} = - H_{\mp}\frac{\partial H_{\pm}}{\partial
t}
\end{equation}
\begin{equation}\label{4eq}
\frac{dp_{i}}{d\tau} = - H_{\mp}\frac{\partial H_{\pm}}{\partial
x^{i}}.
\end{equation}
According to equation (\ref{teq}), $\tau$ can be eliminated in
favor of $t$. The system of equations becomes simply the equations
(\ref{2eq})
and (\ref{4eq})  with $t$ written in place of  $\tau$ and the
factor $H_{\mp}$ cancelled everywhere on the right hand side.
Explicitly,
\begin{equation}\label{null1}
\frac{dx^{i}}{dt} =-\beta^{i} \pm \alpha
\frac{\gamma^{ij}p_{j}}{\sqrt{p_{k}\gamma^{kl}p_{l}}}
\end{equation}
and
\begin{equation}\label{null2}
\frac{dp_{i}}{dt} = - \partial_{i}\left(- \beta^{j}p_{j} \pm
\alpha \sqrt{p_{j}\gamma^{jk}p_{j}}\right).
\end{equation}
The eikonal, corresponding to the Hamiltonian of equation (\ref{6eq}),
\begin{equation}
\partial_{a}Sg^{ab}\partial_{b}S = 0
\end{equation}
can be factored similarly. To do so it is a simple matter of
making the replacements $p_{t} \rightarrow
\partial_{t}S $, $p_{i} \rightarrow \partial_{i}S $ in (\ref{p})
to find the following symmetric hyberbolic partial differential
equation
\begin{equation}\label{PDE}
\partial_{t}S = \beta^{i} \partial_{i} S \pm \alpha\sqrt{\partial_{i}S
\gamma^{ij} \partial_{j}S} \equiv \bar{H}.
\end{equation}
Note that a bar is introduced here to distinguish the Hamiltonian
used here from the Hamiltonian used in (\ref{6eq}). 
This equation is also used in the method of \cite{eh1:}, although
in that work the equation is further reduced to consider the case of 
a single null surface.

The right hand side of this result is homogeneous of degree one in
$\partial_{i}S$.  The characteristic curves along which the level
sets $\Gamma$ of $S$ are propagated, are then
\begin{equation}\label{x2}
\dot{x}^{i} = -\beta^{i} \pm \alpha \frac{\partial^{i}S} {
\sqrt{\partial_{i}S \gamma^{ij} \partial_{j}S}} \equiv
\frac{\partial \bar{H}}{\partial\left(\partial_{i} S\right)}
\end{equation}
\begin{equation}\label{p2}
\partial_{i}\dot{S} = \partial_{i} \bar{H}\left(t,x^{j},\partial_{j}S\right),
\end{equation}
which are the null geodesics of equations (\ref{null1}) and (\ref{null2}). 
Immediately, the integral curves of
the gradients of $S$ and $\Gamma$ are also the null geodesics:
\begin{equation}
\frac{dx^{i}\left(\lambda\right)}{d\lambda} = g^{i a} p_{a}  = g^{i a} 
\partial_{a} S = \partial^{i}
S\left(\lambda, x^{j}\left(\lambda\right)\right).
\end{equation}

Hereafter, the bar on $\bar{H}$
is dropped with the understanding that the Hamiltonian considered is 
that of equation (\ref{PDE}).
This result establishes that the eikonal is technically a Riemann
invariant of the null geodesics, a fact that proves useful
in establishing the signature of a black hole event horizon
in solutions of the eikonal equation. More specifically, since 
the eikonal is the canonical generator of null geodesics, it
can be employed in
analysis of black hole event horizons, which are 
by definition generated by null
geodesics having no future end points. To proceed in this
manner the equation (\ref{PDE}) is employed in an initial value
problem and then surveyed for signature features of black hole
event horizons.

A black hole event horizon is generated by 
a congruence of outgoing --- but future asymptotically nonexpanding
--- null geodesics. The scope of the surveys of the eikonal equation
that are  required to identify a black hole 
event horizon is then restricted to 
the space of all outgoing null surfaces. These surveys are
greatly reduced by the fact that solutions of the eikonal are
categorized by topologically equivalent solutions. To see this,
note that
\begin{equation}
\partial_{i}\psi\left(S\right) = \frac{\partial \psi}{\partial S}
\partial_{i} S = \lambda\left(S\right) \partial_{i}S,
\end{equation}
and
\begin{equation}
\partial_{t}\psi\left(S\right) = \frac{\partial \psi}{\partial S}
\partial_{t} S = \lambda\left(S\right) \partial_{t}S.
\end{equation}
By homogeneity of the right hand side of (\ref{PDE}), if $S$ is a
solution then $\psi\left(S\right)$ is also solution. Thus smoothly
related initial data $S_{0} \rightarrow S_{0}' =
\psi\left(S_{0}\right)$ have smoothly related solutions. This
feature alleviates the need for surveying over smoothly related
initial data.

A further reduction of the scope of solution surveys is provided
by the equivalence of ingoing and outgoing solutions under time
reversal. Propagation of data for $S$ describing an ingoing or outgoing 
null surface is
accomplished by specification of: 1. A definition of the direction
of time, 2. A choice of $\alpha$ and $\beta^{i}$, and 3. A choice
of the root. With these choices specified, data is then uniquely
partitioned into an ingoing type and an outgoing type with the
distinction being the gradient of $S$.

With the dynamics of any outgoing null surface (including the
event horizon) specified by (\ref{PDE}), and the scope of solution
surveys categorized into topological classes, the task remains of
identifying the event horizon within this restricted space of 
outgoing null surfaces. To do so, the results and approach
of \cite{eh1:} are adopted here 
and
modified to include the eikonal equation as discussed above. The
result of this approach is a signature feature of black hole event
horizons in the eikonal equation.

An event horizon of
a black hole is by definition a critical outgoing surface when
tracked into the future. Let $\mathcal{P}$ be a
point interior to the horizon and $\mathcal{Q}$ be a point
exterior to the horizon such that $\mathcal{P}$ and $\mathcal{Q}$
lie on characteristic curves of the eikonal $\gamma_{{\mathcal{P}}}$ and
$\gamma_{{\mathcal{Q}}}$. At arbitrarily early times let
$\gamma_{{\mathcal{P}}}$ and $\gamma_{{\mathcal{Q}}}$ pass
arbitrarily closely to a point $\mathcal{H}$ that lies on the
horizon $\Gamma$. Since $S$ is a Riemann invariant, at arbitrarily
early times the jump of the eikonal at $\mathcal{H}$ becomes
$\left[\left[S\left(\mathcal{H}\right)\right]\right] \equiv
S_{-}\left(\mathcal{H}\right) - S_{+}\left(\mathcal{H}\right)
\approx S\left(\mathcal{P}\right) - S\left(\mathcal{Q}\right) \neq
0$. In the computational domain, where the resolution is finite,
this discontinuity will appear generically in finite time. As
such, an approximation of the  event horizon will appear
numerically as the formation of a jump discontinuity in the
eikonal for outgoing data that is propagated into the past.  This
is the numerical signature of black hole event horizons in the
eikonal's solutions.

\section{Numerical Methods}
Analysis of the continuum properties of the eikonal equation as a
Hamilton Jacobi equation identifies
three closely related approaches to tracking black hole event
horizons:
\begin{itemize}
\item System I: The null geodesic equations
\begin{equation}
S\left(x^{j},t\right) = S_{0}\left(x^{k}\right) - \int dt
H\left(t, x^{i}, p_{i}\right)
\end{equation}
where the integral is evaluated along the solutions of
\begin{eqnarray}
\dot{x}^{i} &=& \beta^{i} \pm \alpha
\frac{p^{i}}{\sqrt{p_{i}\gamma^{ij}p_{j}}},\\
\nonumber \dot{p}_{i} &=& -\partial_{i}H. \\ \nonumber
\end{eqnarray}

\item System II: The eikonal
\begin{equation}\label{eikonal}
\partial_{t} S = -H\left(t, x^{j}, \partial_{j}S\right)
\end{equation}

\item System III: The flux conservative form
\begin{equation}
S\left(x^{i},t\right) = \oint dx^{i}  p_{i}\left(x^{j},t\right)
\end{equation}
\begin{equation}\label{flux}
\partial_{t} p_{i}\left(x^{i},t\right)
+ \partial_{i}H\left(t, x^{j}, p_{j}\right) = 0.
\end{equation}
\end{itemize}
In each of the above three systems of equations the Hamiltonian is
given by (\ref{PDE}).

In each of the above cases, the structure of the dynamical
equations provide certain advantages. For example, in each case the
symplectic structure can be employed to identify 
numerical loss of accuracy in a similar manner to some modern
numerical schemes used in Hamiltonian dynamics. 
Further, as a flux conservative system,
high resolution methods from computational fluid dynamics can be
applied directly to the third system. In following sections of this
article, the
second system is considered since this system of equations 
yields an expedient implementation that is sufficiently accurate
for our purposes.

Singular behavior on the eikonal is not specific to only the event
horizon and instead, as described in detail by Arnold and Newman
\cite{newman1:}, \cite{newman2:}, \cite{newman3:},
the eikonal is known to generically break down on caustic and
other sets. Special numerical methods are then required to handle
the generic singular behavior of the eikonal; we make use of an explicit viscosity term.  In the
continuum, addition of our form of numerical viscosity at the level of the
finite difference approximation corresponds to replacing the
evolution of
\begin{equation}
\partial_{t} S = - H\left(t,x^{i},\partial_{j}S\right)
\end{equation}
with evolution of the equation
\begin{equation}\label{WKB}
\partial_{t} \psi =
\epsilon^{2} \nabla^{2} \psi -
H\left(t,x^{i},\partial_{j}\psi\right)
\end{equation}
where $\epsilon$ is a small quantity which we call the viscosity
and $\nabla^{2}$ denotes any
second order, linear derivative operator. There is a well defined
sense in which the solutions $S$ relate to the solutions $\psi$;
it is simply given (when the solutions $S$ exist) by the WKB
transformation
\begin{equation}
\psi\left(x^{i},t\right) = \sum_{n}
a_{n}\left(x^{i},t\right)\epsilon^{n}
\exp\left(S/\epsilon\right)\equiv A \exp(S/\epsilon).
\end{equation}
To see the explicit relationship between  the solutions $\psi$ and
the solutions $S$ note that
\begin{equation}
\partial_{t} \psi =
\frac{\psi}{\epsilon}\left(\partial_{t} S + \epsilon
\partial_{t} \log A\right),
\end{equation}
\begin{equation}
\partial_{i} \psi =
\frac{\psi}{\epsilon}\left(\partial_{i} S + \epsilon
\partial_{i} \log A\right),
\end{equation}
\begin{equation}
\nabla^{2} \psi = \frac{\psi}{\epsilon^{2}} \left( \nabla S +
\epsilon \nabla \log A \right)^{2} +
\frac{\psi}{\epsilon}\left(\nabla^{2}S  + \epsilon \nabla^{2} \log
A\right).
\end{equation}
Assuming that the Hamiltonian is homogeneous of degree one in
momentum, and making use of perturbation theory:
\begin{equation}
H\left(x^{i},\nabla_{j}\psi\right) = \frac{\psi}{\epsilon}\left(
H\left(x^{i},\nabla_{j}S\right) + \epsilon
H_{1}\left(t,x^{i},\partial_{j}\log A\right)\right).
\end{equation}
Substituting these results into (\ref{WKB}) and cancelling an
overall factor of $\psi/\epsilon$ gives at lowest, and first
order:
\begin{equation}
\partial_{t}S  =
- H\left(t,x^{i},\partial_{j}S\right).
\end{equation}
\begin{eqnarray}
\epsilon \partial_{t} \log A  &=& - \epsilon
H_{1}\left(t,x^{i},\partial_{j}\log A\right)\\ \nonumber &+&
\epsilon \left(\nabla S + \epsilon \nabla \log A \right)^{2} +
\ldots \\ \nonumber
\end{eqnarray}
Here $H_{1}$ is the first order linear Hamiltonian obtained from
perturbation theory. At zeroth order, $S$ then satisfies the
eikonal equation, while the first order correction is the linear
result of first order perturbation theory and expresses the
evolution of $A$. Higher order results can be found similarly.
Again, these results hold only where solutions of $S$ exist; that
is, away from discontinuities or other solution singularities in
$S$ and its derivatives.

The above analysis is a slight modification of usual WKB expansion
and expresses the solutions $\psi$ in terms of the solutions $S$.
Such solutions can be inverted using the conventional method of
series inversion. To invert the WKB expression, and express the
solutions of the eikonal in terms of the solution of the parabolic
equation, assume first that the solution of the parabolic equation 
(\ref{WKB})
is given as $\psi$ (numerically or otherwise). Writing
\begin{equation}\label{spsi}
S = \log \left(\frac{\psi}{\bar{A}}\right),
\end{equation}
the procedure is then analogous to that of the WKB expansion:
$\bar{A}$ is constructed as an asymptotic power series with
coefficients depending on $\psi$ alone. The result of such an
analysis is simply:
\begin{equation}
S\left(x,t\right) = \log \psi\left(x,t\right)  + \bar{\epsilon}
\int dt' \frac{\nabla^{2}\psi\left(\bar{x},t\right)}{\psi}
\end{equation}
where the integral is evaluated along
\begin{equation}
\frac{d\bar{x}^{i}}{dt} = \frac{\partial
H\left(t,\bar{x}^{j},\partial_{\bar{x}^{k}}\log\psi\right)}{\partial
p_{i}}
\end{equation}
and $\psi$ is provided independently by (numerical) solution of
\begin{equation}
\partial_{t} \psi = \bar{\epsilon} \nabla^{2}\psi -
H\left(t,x^{i},\partial_{j}\psi\right).
\end{equation}

One advantage of an explicit second order viscosity term is a
simple procedure for reducing the error of viscosity solutions by one
order of the viscosity $\epsilon$.  The limit $\epsilon \rightarrow 0$ corresponds to a continuous
family of zeroth order solutions $\phi_{\epsilon}$; 
where 
\begin{equation}
\phi_{\epsilon} = S + \eta_{\epsilon}
\end{equation}
and $\eta_{\epsilon} = {\mathcal{O}}\left(\epsilon\right)$. Note that
each solution $\phi_{\epsilon}$ is obtained from an analysis similar
to that following equation (\ref{spsi}), although obtained
by neglecting the logarithm. 

Given two viscosity solutions $\phi_{\epsilon}$ and $\phi_{2
\epsilon}$ it is then possible to construct a third improved
viscosity solution $\phi_{I}$ that is accurate to
${\mathcal{O}}\left(\epsilon^{2}\right)$. To see this, consider the
combination
\begin{equation}
\phi_{I} \equiv \phi_{\epsilon}  + \left(\phi_{\epsilon} - \phi_{2
\epsilon}\right).
\end{equation}
$\phi_{I}$ is accurate to
${\mathcal{O}}\left(\epsilon^{2}\right)$ since
\begin{equation}
\phi_{\epsilon} - \phi_{2 \epsilon} = \eta_{\epsilon} - \eta_{2
\epsilon} \approx  \eta_{\epsilon} - 2 \eta_{\epsilon} = -
\eta_{\epsilon}.
\end{equation}

To make use of these improved viscosity solutions  let $h$ denote
the resolution of the numerical mesh. Any second order finite
difference approximation will have then
\begin{equation}
\phi = \hat{\phi} + {\mathcal{O}}\left(h^{2}\right),
\end{equation}
where $\phi$ is the continuum solution and $\hat{\phi}$ denotes
its finite difference approximation.[We will use a hat  throughout
the paper to denote the discrete approximation to a continuum object.] Similarly,
\begin{equation}
S = \hat{S} + {\mathcal{O}}\left(h^{2}\right).
\end{equation}
Using
\begin{equation}
S =\phi + {\mathcal{O}}\left(\epsilon\right) =  \phi_{I} +
{\mathcal{O}}\left(\epsilon^{2}\right)
\end{equation}
gives
\begin{equation}
\hat{S}= \hat{\phi} + { \mathcal{O}}\left(h^{2}\right) +
{\mathcal{O}}\left(\epsilon\right) = \hat{\phi_{I}} + {
\mathcal{O}}\left(h^{2}\right) +
{\mathcal{O}}\left(\epsilon^{2}\right).
\end{equation}

\section{Level Set Extraction}
Of crucial interest in the binary black hole coalescence problem
are the areas of sections of the black hole event horizon.
To find these sections, at any given time level  
any of the level sets, say $S =
0$, can be extracted to obtain the surface
$\Gamma$, a two dimensional section of the corresponding null surface. 
This problem of extraction is an inverse problem, since
it requires that points $\left(x,y,z\right)$ are found such that
$S\left(x,y,z\right) = 0$. To accomplish this inversion, we find that
an ordinary
bisection method is sufficient for use with an ordinary second order
interpolation scheme. In this method it is assumed that the
surface $\Gamma$ can be expressed in spherical coordinates
$\left(\theta, \phi, u\left(\theta, \phi\right)\right)$, where $r
= u\left(\theta, \phi\right)$ is the surface function for a given
center $c^{i}$ contained within the surface $\Gamma$. Given a
choice for the center, the radial function for
$S\left(\Gamma\right) = S_{o} = 0$ can then be approximated via the interpolation
and bisection.

To establish the accuracy of our implementation, including
the routines that accomplish 
 extraction of  the level sets and the accuracy of 
the viscosity term, we consider stationary, spinning black holes. 
This case is
completely described by the Kerr - Newmann family of axisymmetric solutions of Einstein's equation. It is convenient to make
use of the Kerr-Schild form for the metric \cite{deirdre:} since 
this form is used in our binary black hole evolution code as well
as in our solution of the initial value problem for setting
initial data for the evolution. Specifically,
the metric in the Kerr-Schild form is
\begin{equation}
g^{ab} = \eta^{ab} - 2 H l^{a}l^{b}.
\end{equation}
Here $\eta_{ab} = \mathrm{diag}\left(-1,1,1,1\right)$ is
Minkowski's metric, $H$ is a space time scalar, and $l^{a}$ is an
ingoing null vector with respect to both the Minkowski and full
metric. The Kerr solution is the two parameter family of solutions
such that
\begin{equation}
H = \frac{M r^{3}}{r^{2} + a^{2} z^{2}}
\end{equation}
and
\begin{equation}
l^{t} = -1,
\end{equation}
\begin{equation}
l^{x} = \frac{r x + a y}{r^{2} + a^{2}},
\end{equation}
\begin{equation}
l^{y} = \frac{r y - a x}{r^{2} + a^{2}},
\end{equation}
\begin{equation}
l^{z} = \frac{z}{r},
\end{equation}
\begin{equation}
r^2 = \frac{1}{2}\left(\rho^{2} - a^{2}\right) +
\sqrt{\frac{1}{4}\left(\rho^{2} - a^{2}\right) + a^{2} z^{2}}
\end{equation}
where
\begin{equation}
\rho^2 = x^{2} + y^{2} + z^{2}.
\end{equation}
Finally, the event horizon for the Kerr black hole is located
on the ellipsoid $r = r_{+} = r\left(x,y,z\right)$ where
\begin{equation}
r_{+} = M + \sqrt{M - a}.
\end{equation}
\begin{figure} 
\epsfxsize=8cm
\centerline{\epsfbox{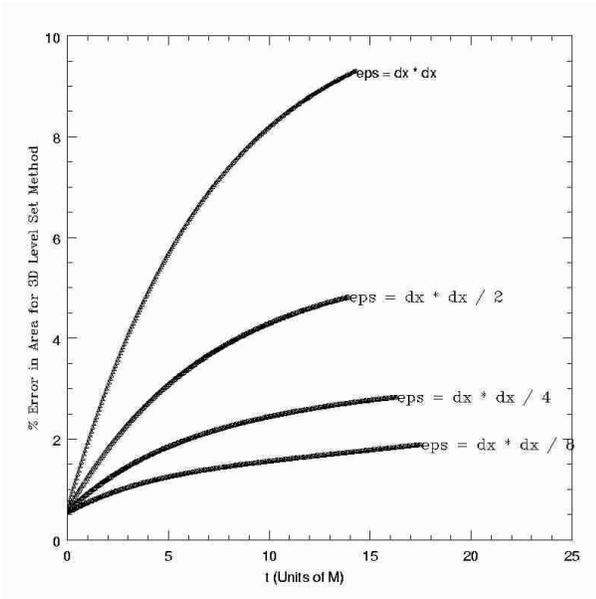}}
\vspace{0.5cm}
\caption{Percent error in area of $M = 2$, $a = 0$ event
horizons in survey over viscosity parameter: $\epsilon = $
$h^2$, ${h^2}/2$, ${h^2}/4$, ${h^2}/8$. Here increasing $t$ corresponds
to propagation into the past.} \label{verrora:1}
\end{figure}

\begin{figure} 
\epsfxsize=8cm
\centerline{\epsfbox{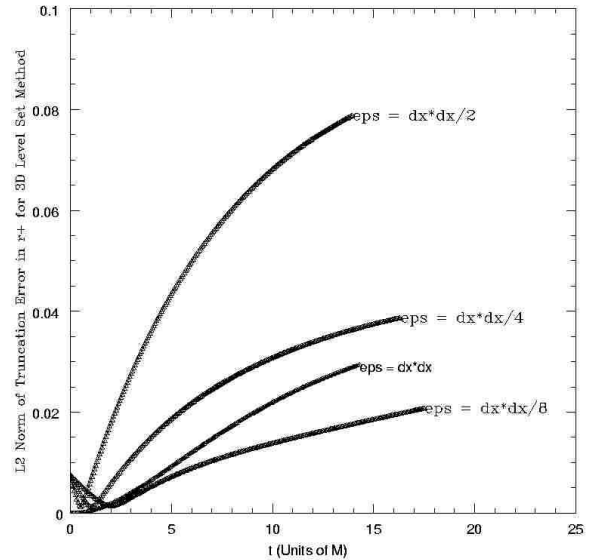}}
\vspace{0.5cm}
\caption{L2 norm in truncation error of $r_{+}$ for $M = 2$, $a = 0$ event
horizons in survey over viscosity parameter: $\epsilon = $
$h^2$, ${h^2}/2$, ${h^2}/4$, ${h^2}/8$. Here increasing $t$ corresponds
to propagation into the past.} \label{vrmse:1}
\end{figure}

\begin{figure} 
\epsfxsize=8cm
\centerline{\epsfbox{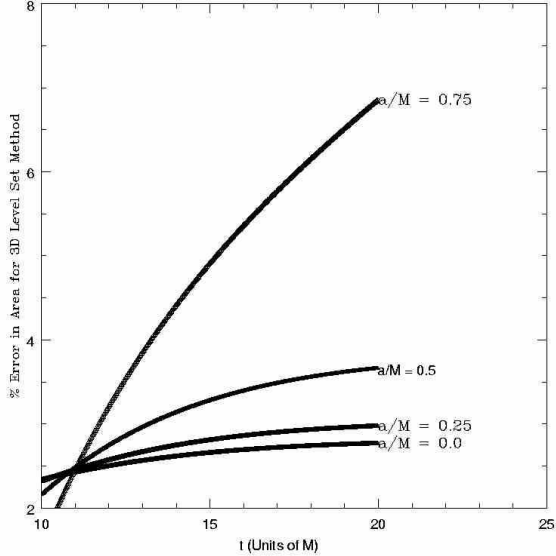}}
\vspace{0.5cm}
\caption{Percent error in area for three dimensional
level set solutions $a/M =$, $0$, $1/4$, $1/2$, $3/4$. Here 
increasing $t$ corresponds
to propagation into the past.} \label{vserrora:1}
\end{figure}

\begin{figure} 
\epsfxsize=8cm
\centerline{\epsfbox{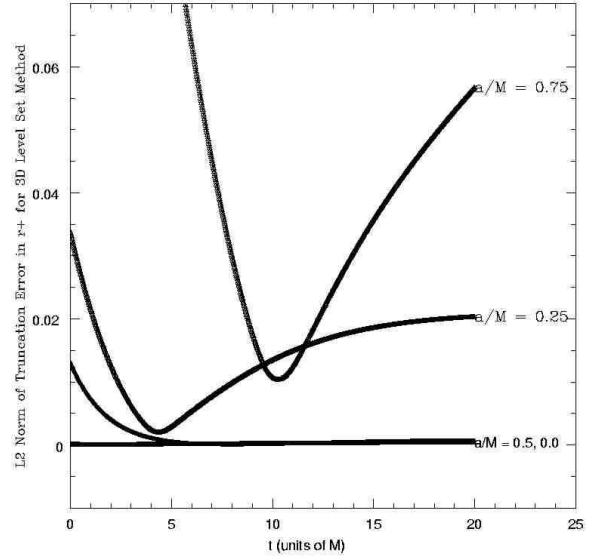}}
\vspace{0.5cm}
\caption{L2 norm of truncation error in $r_{+}$ for three dimensional
level set solutions $a/M =$ $0$, $1/4$, $1/2$, $3/4$. Here increasing $t$ corresponds
to propagation into the past.} \label{vsrmse:1}
\end{figure}

In figures (\ref{verrora:1}) -(\ref {vsrmse:1}) we show the evolution 
backward in time of the eikonal equation (followed by extraction of the $S=0$ 
surface). Errors can arise both in the evolution and in the extraction of the 
surface. Figure (\ref{verrora:1}) shows the percent errors in the extracted 
areas for a nonspinning black holes with mass $M = 2$, $a = 0$ in a survey over 
the viscosity parameter. Figure (\ref{vrmse:1}) shows a similar study but for 
the L2 norm of the truncation error in the function $r_{+}$. Note that 
the function $r_{+}$ is defined for every point of the horizon in the
continuum $r_{+} = r_{+}\left(x,y,z\right)$. In the computational domain
$r_{+}$ then takes a discrete form 
$\hat{r}_{+} = \hat{r}_{+}\left(i,j,k\right)$, where the integers $i,j,k$ span
the numerical mesh. The truncation error $e_{r_{+}}$ is then
\begin{equation}
e_{r_{+}}\left(i,j,k\right) = r_{+}\left(x,y,z\right) - \hat{r}_{+}\left(i,j,k\right)
\end{equation}
where is it understood that both $r_{+}$ and $\hat{r}_{+}$ are evaluated
at the same point. Bith Figures(\ref{verrora:1}) and (\ref{vrmse:1})show that 
viscosity parameter $h^2/8$ adequately captures the horizon location. While not 
perfect, it will suffice for the short term horizon tracking reported
here.
However these results also suggest that in vanishing viscosity the percentage error in
the calculated area is reduced toward a bias. This bias is partly
associated to the finite resolution of the computational mesh and
partly to accuracy of the extraction routine. These figures were
generated using a three dimensional computational domain of
$N^{3}$ points with $N = 121$. The outer boundaries are located at
$\left[-15M, +15M\right]$ in the $x,y,z$ directions. The
resolution of the finite difference mesh for these results is then
$h = M / 4$. Also, a Courant -  Friedrichs -  Lewy factor of
$\lambda = 1/4$ with an iterated Crank Nicholson scheme \cite{icn:}
was used
as the finite difference approximation of the evolution of the eikonal
equation (\ref{PDE}).
Neumann boundary conditions $\partial_{i} S = 0$ on the outer boundary 
are found
to be generically
sufficient conditions for stability of the method. The philosophy
here is that the primary interest is deep within the bulk of the
computational domain where the event horizon of the black hole is
located. The outer boundary is then treated only to the degree that the
evolution of the interior region remains stable. Further, the
interior of the black hole is excised from the computational
domain in a sphere of radius $r = r_{0} + 2 dx$, where $r_{0}$
denotes the radius of the Kerr - Newman ring curvature
singularity. Finally, the discrete surface $\hat{\Gamma}$ was
constructed using $m^{2}$ points with $m = 100$. At such a
resolution of the extracted surface any errors in the area must be
attributed to all of: the viscosity parameter, the extraction routine, and
the resolution of the underlying three dimensional grid.

Figures (\ref{vserrora:1}) and (\ref{vsrmse:1}) show the percent
 errors in the extracted areas
as well as the L2 Norm of the truncation error in the function
$r_{+}$ in a survey over the angular momentum parameter $a$.
Here the viscosity is $\epsilon = h^{2}$. 
These figures both show the
evolution from a single null sphere that is completely
exterior to the horizon and propagated into the past. 
Since the event horizon of a spinning black hole 
is elliptical in its geometry, the spherical data we 
have chosen for $t_{\downarrow} = 0$ corresponds 
to a percent error in the area and $r_{+}$ that varies 
with the spin parameter $a/M$ at 
$t_{\downarrow} = 0$, explaining why the curves do not 
intersect at $t_{\downarrow} = 0$. (Where appropriate we append 
a $\downarrow$ to $t$, thus: $t_{\downarrow}$, 
indicating evolution into the past;
we also sometimes use $t_{\uparrow}$ to emphasize that we mean 
the forward evolving, usual, time $t$. Thus 
$t_{\downarrow} = 0$ corresponds to the late time at which we begin to 
integrate into the past.) 

However, the errors should converge to a constant, which
is evident in each of the curves with $a<0.75$ in 
figures (\ref{vserrora:1}) and (\ref{vsrmse:1}). For $a=0.75$ the curve
does not converge, and we expect that different
choice of initial data {\it will} exhibit convergence. 
According to these results, the viscosity level set method does
indeed accurately and robustly detect the distorted outermost
event horizons of spinning black holes at least when $a<0.75$. 
Note that we study both
the accuracy of $r_{+}$ and the accuracy of the calculated
areas since the calculations seperately and together establish
the accuracy of our area calculation and of our detection of the 
event horizon. 

\section{Asymmetric Binary Black Hole Coalescence}
Analysis of the event horizon for the binary black hole
coalescence problem in  the case of head on collision has been
considered in detail in \cite{eh4:}. The problem of the event
horizon for asymmetric, that is off axis, collision has only been
considered analytically \cite{wini2:} and prior to this work no
results for numerically generated sources have been analyzed.
Numerical evolution and analysis of an asymmetric  binary
black hole system was studied in \cite{exc1:}, but
at that time the question of the event horizon was not considered.

In this section the results of the previous sections are applied
to the first completely numerical analysis of the event horizon
for the case of asymmetric collision.

To begin, consider two black holes of mass $M = 1$ with aligned
spins in the positive $z$ direction of $a/M = 1/2$. The 
computational domain is a grid of
$N^{3}$ points with $N = 121$. The outer boundary is located at
$\pm 15 M$ and the holes are initially positioned at 
$\left(x,y,z\right) = \left(+6,+2,0\right)$, and 
$\left(x,y,z\right) = \left(-6,-2,0\right)$. This computational domain
is identical to the mesh used in  the previous section to analyze
the percent error in area calculations  of surfaces extracted from
the level set method. The percent error in the
calculation of the area of sections of the horizon should
have a magnitude of about $4-5\%$. Further, as an
order of magnitude estimate, in a flat spatial geometry; e.g., in
a Newtonian spacetime, the initial separation $s$ of the black
hole centers would be $s = \sqrt{12^2 + 4^2} \approx 12.64$. This
would seem to be an ample initial separation to guarantee that the
initial data corresponds to two distinct black holes. However, in
the case that the holes are nonspinning, each will have a
spherical event horizon of radius $r = 2M$. Assuming only marginal
distortion of the nonspinning event horizons due to spin effects,
which is an approximation that is justified by the properties of spin
$a/M= 1/2$ black holes, the nearest separation
between the two sections of the black hole event horizon
is then
approximately $s_{min} = 12.64 M  - 4 M  \approx 8.64M $. Again,
this approximation assumes a flat underlying geometry; and so can
be considered only as an order of magnitude estimate. These initial
data then appear to correspond to a separation of approximately two
nonspinning black hole diameters between the surfaces of section
of the black holes. 

The holes are boosted along the $x$
direction with speeds of $\pm c / 2$. This boost lengthens the
nearest separation $s_{min}$ of the holes due to Lorentz
contraction of the horizons.[In this coordinate system the horizons
undergo contraction in the direction of motion. For a single hole, 
the area of the horizon does not change under this boost.] The nearest 
coordinate separation $s_{min}$
between the holes is then expected to lie in the range $8.64 <
s_{min} < 12.64$.

The numerical evolution of this collision process is carried out
for approximately $10M$ of run time with a Courant factor of
$\lambda = dt / dx = 1/4$. The code is the Texas black hole 
evolution code, a derivative of the Agave code \cite{exc1:}. 
Apparent horizon finders
\cite{deirdre:} locate two distinct apparent horizons of area $A
\approx 50 M $ for the initial data and continue to do so until $t
= 8M $, when only a single apparent horizon of area $A \approx
200M$ can be located. This single apparent horizon persists until
approximately $10M$, beyond which instability effects, stemming
from the outer boundary and the excision boundary, swamp the
solution.

\section{Data For the Eikonal}
While the run length of this asymmetric collision data is
relatively small in units of the black hole masses, the problem of
detecting the associated black hole event horizon, or horizons, is
not a small computational problem. For example, to analyze this data requires analysis of the
lapse $\alpha$, the shift vector $\beta^{i}$, and the three metric
$\gamma^{ij}$. By symmetry of the metric, $\gamma^{ij} =
\gamma^{ji}$, there are only 6 independent components. Tracking
the event horizon of the associated data then requires analysis of
10 grid functions, where each grid function consists of
${\mathcal{O}}\left(Nt\times N^{3}\right)\times 8 B  =
{\mathcal{O}}\left(160\times121^{3}\right) \times 8 B \approx 2.26
GB$ of data. That is, tracking the associated event horizon
requires analysis of approximately $20 GB$ of data.

A more serious difficulty associated to this data set is the
relaxation time, $t\approx 4 M $, that is typically required for outgoing
data to converge onto the event horizon when followed into the
past. Assuming that the collision time is (as suggested by the
apparent horizon solvers) near $t = 8M$, perturbation theory
implies that the resulting horizon should undergo quasi normal
ringing for another $t \approx 20M$ from that time. That is, at
the time level $t = 10M$, where the event horizon tracker will
begin tracking into the past, it is expected that the event
horizon remains highly distorted and far from its stationary
regime. The problem is to determine good data for the eikonal at
the time level $t = 10M$, which can then be propagated into the
past. In contrast, in analysis of the event horizon for the case
of head on collision, researchers made use of approximately $100M$
of data \cite{eh4:}, and assumed that the final state at $t = 100M$
was a
stationary or quasi stationary black hole. In such a circumstance
using spherically symmetric data at $t = 100M$ for an event
horizon tracking method should prove sufficient.  The difficulty
with a data set only of length $t = 10M$ is that using spherically
symmetric data for the eikonal at the time level $t = 10M$ does
not accurately approximate sections of the event horizon at that
time.

To accommodate data of this $10M$ length and to generate
sufficient data for the eikonal at the time level $t = 10M$,  a
method for finding a candidate section of the event horizon at $t
= 10M$ is proposed here and applied to the problem. This method
makes use of the analytic properties of apparent and event
horizons.

To motivate the method, note that if the event horizon were
stationary at $t = 10M$ then the expansion of the surface would be
vanishing there:
\begin{equation}
\theta\left(t = 10M\right) = \frac{1}{A}\frac{dA}{dt} = 0.
\end{equation}
In such a circumstance the apparent horizon could be used
as initial data for the eikonal, which could
then be propagated into the past. However, according to the second
law of black hole mechanics, in the case that the horizon is 
nonstationary, which is the situation expected for this asymmetric
problem, at $t = 10M$ the horizon will satisfy
\begin{equation}
\theta \left(t = 10M\right)\geq 0,
\end{equation}
in the forward time direction. In the backwards time direction
these dynamics correspond to $\theta \leq 0$ at $t = 10M$. As
such, the Taylor series of any compact null surface,
including the critical surface of the event horizon, is at least
of the form
\begin{equation}
A\left(t + dt\right) = A\left(t\right) + dt A \theta + dt^{2}
\frac{d^{2}A}{2dt^{2}} + \ldots
\end{equation}
and probably cannot be truncated to below
\begin{equation}\label{adot}
A\left(t + dt\right) \approx A\left(t\right) + dt A \theta +
\ldots
\end{equation}
The apparent horizon is then a poor estimate for the
the event horizon at $t = 10M$. The objective is to use this 
behavior (\ref{adot}) in combination with the
property that outgoing null data followed into the past converges
to the event horizon. The hope is to establish a better
approximation for the structure of the section of the event
horizon at $t = 10M$, which can then be used in an event horizon
tracking method.

To proceed, consider three compact null surfaces of outgoing data:
One completely interior to the horizon $\Gamma^{i}$, one
completely exterior to the horizon $\Gamma^{e}$, and one that is a
surface of  section of the event horizon $\Gamma^{h}$. Let
$t_{\downarrow} \rightarrow \infty$ denote propagation into the
past. By the property that outgoing data numerically converges to
the horizon when propagated into the past, $\lim_{t_{\downarrow}
\rightarrow \infty} \Gamma^{i} = \lim_{t_{\downarrow} \rightarrow
\infty} \Gamma^{e} =  \lim_{t_{\downarrow} \rightarrow \infty}
\Gamma^{h}$. Thus in most cases, if the surface
$\Gamma^{i}\left(t_{\downarrow} + dt_{\downarrow}\right)$ is
pulled back to the time level $t_{\downarrow}$ and compared to the
surface $\Gamma^{i}\left(t_{\downarrow} \right)$ it will be
completely exterior to $\Gamma^{i}\left(t_{\downarrow} \right)$.
Similarly, if the surface $\Gamma^{e}\left(t_{\downarrow} +
dt_{\downarrow}\right)$ is pulled back to the time level
$t_{\downarrow}$ it will be completely interior to the surface
$\Gamma^{e}\left(t_{\downarrow} \right)$.  That is, by iteratively
pulling the surfaces back to a single time level, spherically
symmetric data will approach the numerical event horizon.
Exceptions to this general behavior stem from the presence of
caustics in the spacetime, where null surfaces intersect, and in
neighborhoods of the event horizon. In particular, as discussed
above, the horizon of this asymmetric collision data is expected
to satisfy $\theta \geq 0 $. If the
surface $\Gamma^{h}\left(t_{\downarrow} + dt_{\downarrow}\right)$
is then pulled back to  $t_{\downarrow}$ it will  be interior to
the surface $\Gamma^{h}\left(t_{\downarrow}\right)$. Let
$\Gamma^{i}_{\epsilon}\left(t_{\downarrow}\right)$ be a
perturbation of $\Gamma^{h}\left(t_{\downarrow}\right)$ that is
arbitrarily close to the horizon
$\Gamma^{h}\left(t_{\downarrow}\right)$ but completely interior to
the surface $\Gamma^{h}\left(t_{\downarrow}\right)$. Due to round
off and truncation error of the finite difference approximation
and the fact that $\lim_{t_{\downarrow}\rightarrow \infty}
\Gamma^{i}_{\epsilon} = \Gamma^{h}$, if the numerical finite
difference approximation
$\hat{\Gamma}^{i}_{\epsilon}\left(t_{\downarrow} +
dt_{\downarrow}\right)$, which approximates the continuum section
$\Gamma^{i}_{\epsilon}\left(t_{\downarrow} +
dt_{\downarrow}\right)$, is pulled back to the time level
$t_{\downarrow}$ it will generically intersect
$\hat{\Gamma}^{i}_{\epsilon}\left(t_{\downarrow}\right)$ and
contain neighborhoods that are both interior and exterior to the
finite difference approximation
$\hat{\Gamma}^{i}_{\epsilon}\left(t_{\downarrow}\right)$. Similar
behavior will hold for numerical data
$\hat{\Gamma}^{e}_{\epsilon}\left(t_{\downarrow}\right)$ defined
to be a perturbation of
$\hat{\Gamma}^{h}\left(t_{\downarrow}\right)$ that is arbitrarily
close to the horizon $\hat{\Gamma}^{h}\left(t_{\downarrow}\right)$
, and completely exterior to
$\hat{\Gamma}^{h}\left(t_{\downarrow}\right)$.

According to this behavior, data for the eikonal at the time level
$t = 10M$ can be constructed by iteration over that time slice.
This approach is similar to treating the event horizon as if it
were an apparent horizon, although modified to account for
this nonstationary regime. Beginning with spherically symmetric
initial data that is well exterior to the horizon
$\hat{\Gamma}^{e}_{1}\left(t_{\downarrow} = 10M\right)$ the data
is updated, pulled back to the original time level and reset as
follows: $\hat{\Gamma}^{e}_{2}\left(t_{\downarrow}\right) =
\hat{\Gamma}^{e}_{1}\left(t_{\downarrow} +
dt_{\downarrow}\right)$. The step
$\hat{\Gamma}^{e}_{n+1}\left(t_{\downarrow}\right) =
\hat{\Gamma}^{e}_{n}\left(t_{\downarrow} + dt_{\downarrow}\right)$
is then repeated for several hundred iterations, which corresponds
to ${\mathcal{O}}\left(10\right)$ $e$ folding times. Note that in
the case that the spacetime is stationary the surface
$\hat{\Gamma}$ will converge to the apparent horizon according to
this method. Since the apparent horizon of a stationary spacetime
coincides with the event horizon, this method will generically
find the event horizon in the case of stationary spacetimes.
However, in the nonstationary regime, which is the case for the
asymmetric collision problem, the result of this procedure is at
best an improved initial guess for an event horizon tracker.
Further information, such as area analysis or study of the
apparent horizon, is required to argue that the resulting surface
is a candidate section of the event horizon.

\begin{figure} 
\epsfxsize=8cm
\centerline{\epsfbox{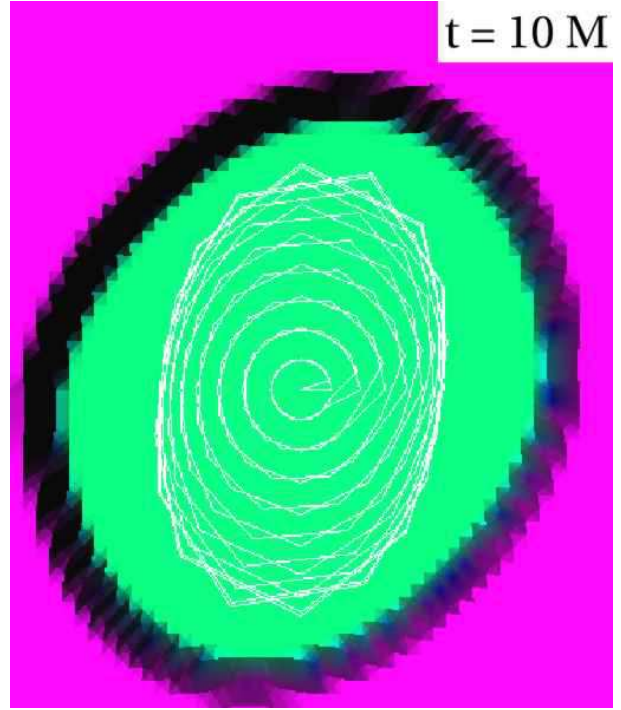}}
\vspace{0.5cm}
\caption{Tanh
data (see Eq(\ref{id}) for the eikonal in asymmetric binary black hole
coalescence. The black contour is the estimate of the event horizon
section, while the wire mesh is the apparent horizon. These data are set at the end 
of the computational evolution ($t=10M$), and will be evolved into the past.} \label{m:1}
\end{figure}

\begin{figure} 
\epsfxsize=8cm
\centerline{\epsfbox{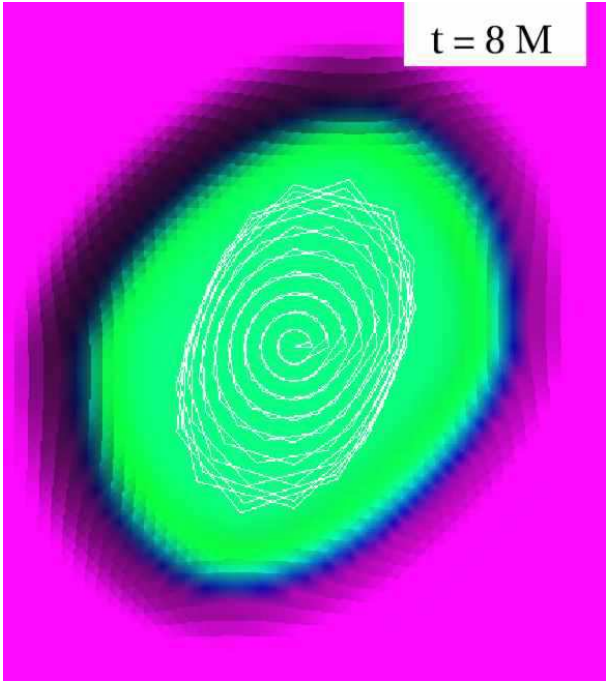}}
\vspace{0.5cm}
\caption{Data of figure (\ref{m:1}) (asymmetric binary black hole
coalescence) evolved backward from $t = 10M $ to $t = 8M$.} \label{m:2}
\end{figure}

We show in figure (\ref{m:1}) the result of such a process, applying
$\approx 200$ iterations over the time slice $t = 10M$. Figures (\ref{m:1})
through (\ref{bh:8}) show the eikonal function in the $z=0$ plane. The location
of the determined guess for the surface $\Gamma$ (which we will take to begin
our evolution of the horizon into the past) is encoded into the eikonal
by the color map. With this $\Gamma$, data for the eikonal can be written
in the form: 
\begin{equation}\label{id}
S\left(0,x^{i}\right) = 1 + \tanh\left(\frac{r_{c} - r}{c}\right)
\end{equation}
In equation (\ref{id}) the first argument of the eikonal is $t_{\downarrow} = 0$;
$t_{\downarrow}$ will increase into the past. Also, $r_{c}$
denotes the data $\Gamma$ and $c$ controls its steepness. Typically
we take a transition width $c$  on the order of a computational zone. Note
that this surface is not considered to be a true section of the
event horizon, but instead is a good initial guess or candidate
section, which after a few $M$ will evolve into better
approximation of a true section of the event horizon. By way of
comparison figure (\ref{m:1}) also shows at $t = 10M$ the final apparent
horizon as a white wire mesh. Note that both the apparent horizon and
these data for the eikonal are highly distorted from the
stationary case. Figure (\ref{m:2}) shows the resulting eikonal
function $S\left(x,y,z\right)$ after $2M$ of evolution into the past.  
Note that
the surface $\Gamma$ is not qualitatively changed during the
evolution.

\section{Surface Extraction and Apparent Horizons}
\begin{figure} 
\epsfxsize=8cm
\centerline{\epsfbox{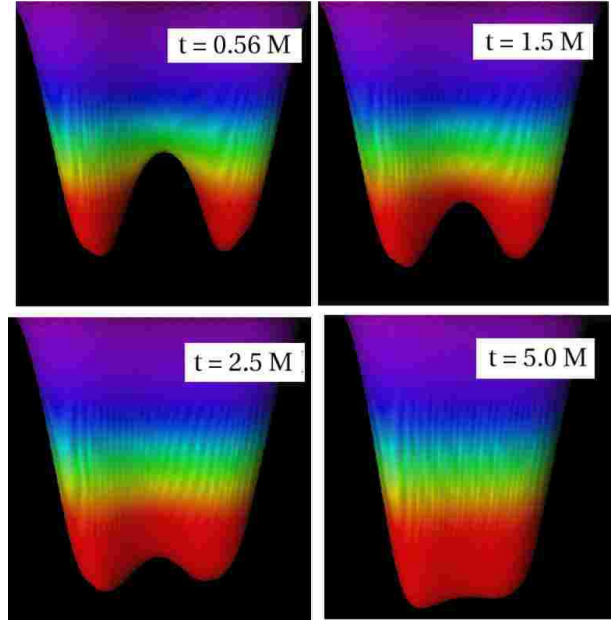}}
\vspace{0.5cm}
\caption{Change of
topology in eikonal for asymmetric binary black hole coalescence, 
shown as an elevation map.}
\label{m:4}
\end{figure}

Figure
(\ref{m:4}) shows several frames of the
evolution of the eikonal using a viscosity solution of $\epsilon =
h^{2}$ (not of the improved viscosity form). This figure
display the value of the eikonal function on the $z=0$ plane. 
Figure (\ref{m:4}) shows the eikonal
data as an elevation map and also via the color map of (\ref{m:1}). 
Note in those figures that null surfaces
interior to the event horizon undergo a change in topology and
this topological transition is continuously monitored by the
viscosity solutions of the eikonal. In figure   (\ref{m:4})
$t = 0.562 M$  is shown in the upper left-hand corner,
$t=1.5M$ is shown in the upper right-hand corner, $t = 2.5M$ is shown 
in the lower
left-hand corner and  $t = 5.0M$ appears in the lower
right-hand corner.


\begin{figure} 
\epsfxsize=8cm
\centerline{\epsfbox{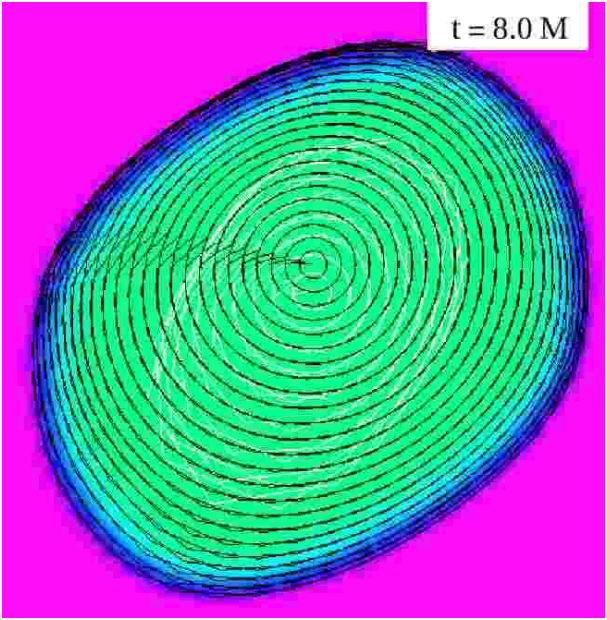}}
\vspace{0.5cm}
\caption{Asymmetric
binary black hole coalescence:  $t_{\uparrow} = 8M$. This is the 
same as figure (\ref{m:2}) but the black wire mesh shows the estimated 
location of the event horizon.} \label{m:5}
\end{figure}

\begin{figure} 
\epsfxsize=8cm
\centerline{\epsfbox{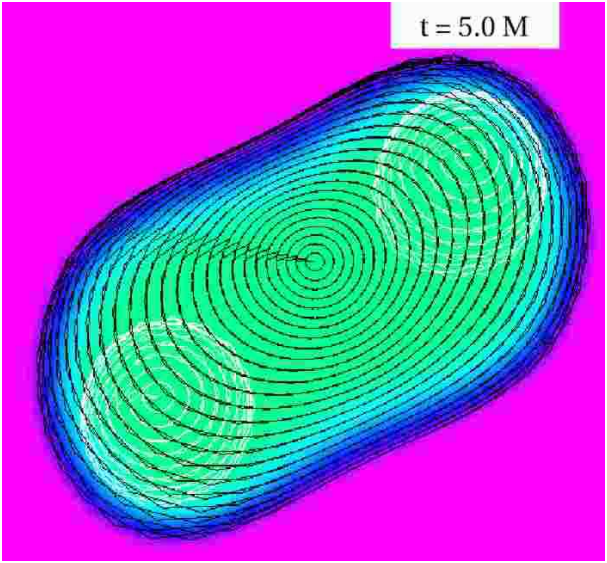}}
\vspace{0.5cm}
\caption{Asymmetric
binary black hole coalescence:  $t_{\uparrow} = 5M$. Note 
that while the apparent horizons (the white wire-frame
``spheres") are still well separate at $t=5M$, 
the event horizon (black wire frame ``peanut" 
already has one component only.} \label{m:6}
\end{figure}

\begin{figure} 
\epsfxsize=8cm
\centerline{\epsfbox{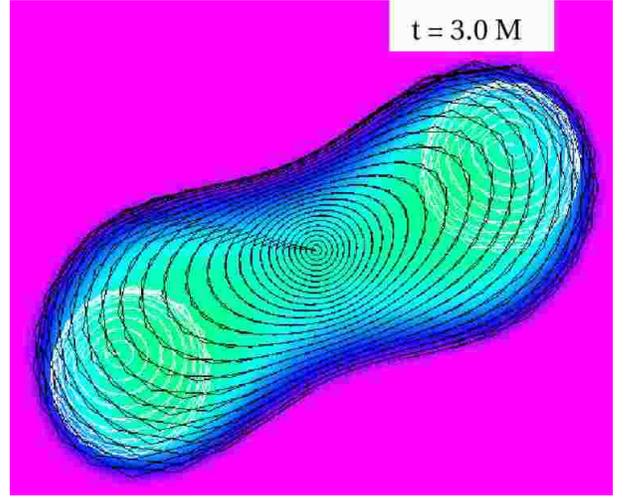}}
\vspace{0.5cm}
\caption{Asymmetric
binary black hole coalescence: $t_{\uparrow} = 3M$.} \label{m:7}
\end{figure}

\begin{figure} 
\epsfxsize=8cm
\centerline{\epsfbox{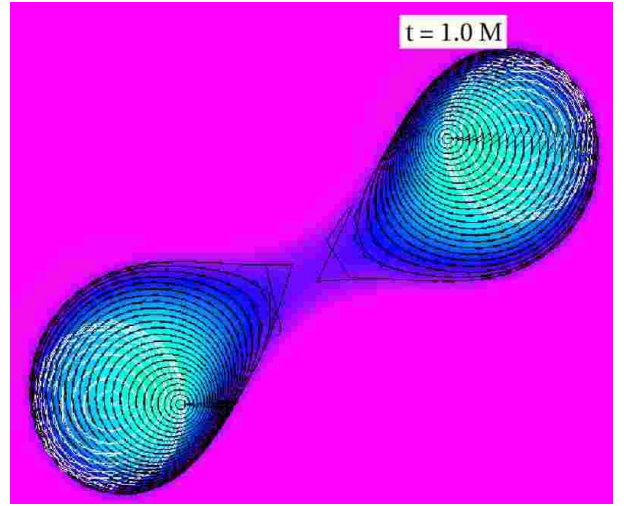}}
\vspace{0.5cm}
\caption{Asymmetric
binary black hole coalescence: $t_{\uparrow} = 1M$.
Careful inspection of the event horizon (black wireframe) suggests two separated 
components.} \label{bh:8}
\end{figure}

The figures (\ref{m:5}) - (\ref{bh:8}) continue the sequence of 
figures (\ref{m:1}) - (\ref{m:2}), and show, for several
values of $t_{\uparrow}$, the value of the eikonal in the $z=0$ plane;
the location of the apparent horizon in the $3$ dimensions (the white
wire frame); and in the black wire frame, locations of sections of a  candidate event horizon
$\hat{\Gamma}_{c}\left(t_{\downarrow}\right)$ that is generated by
evolution of the eikonal equation from data constructed
using the method described in the previous section. In this
context, the surfaces
$\hat{\Gamma}_{c}\left(t_{\downarrow}\right)$ are extracted from
the eikonal data using the technique described in section IV.
Note that
 $\hat{\Gamma}_{c}$ completely
contains the apparent horizons throughout their evolution. This is
a fundamental condition that any numerically constructed black
hole event horizon must satisfy. To determine how these results
depend on the initial data $\hat{\Gamma}_{c}\left(t_{\downarrow} =
0 \right)$, choosing initial data
$\hat{\Gamma}_{\delta}\left(t_{\downarrow} = 0 \right)$ of the
form
\begin{equation}
u_{\delta}\left(\theta, \phi\right) = u_{c}\left(\theta,
\phi\right) - \delta,
\end{equation}
permits survey about the data $u_{c}\left(\theta, \phi\right)$,
where $u_{c}\left(\theta, \phi\right)$ corresponds to the data
$\hat{\Gamma}_{c}\left(t_{\downarrow}= 0\right)$. Studies with
$\delta = M/2, M, 2 M $ establish that the level sets
$\hat{\Gamma}_{\delta}$ penetrate both apparent horizons for any
$\delta \geq M/2$. These results suggest that the true event
horizon is contained in a domain parameterized by  $0 < \delta <
M/2 $.

\begin{figure} 
\epsfxsize=8cm
\centerline{\epsfbox{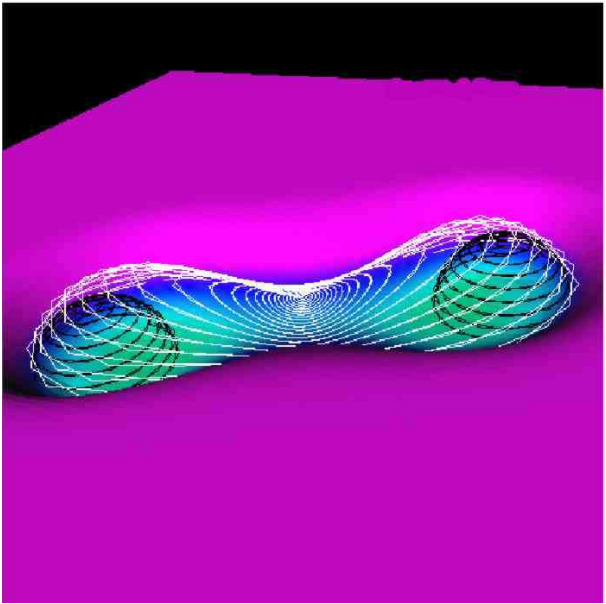}}
\vspace{0.5cm}
\caption{Level
set extraction for asymmetric binary black hole coalescence: I.}
\label{m:8}
\end{figure}

\begin{figure} 
\epsfxsize=8cm
\centerline{\epsfbox{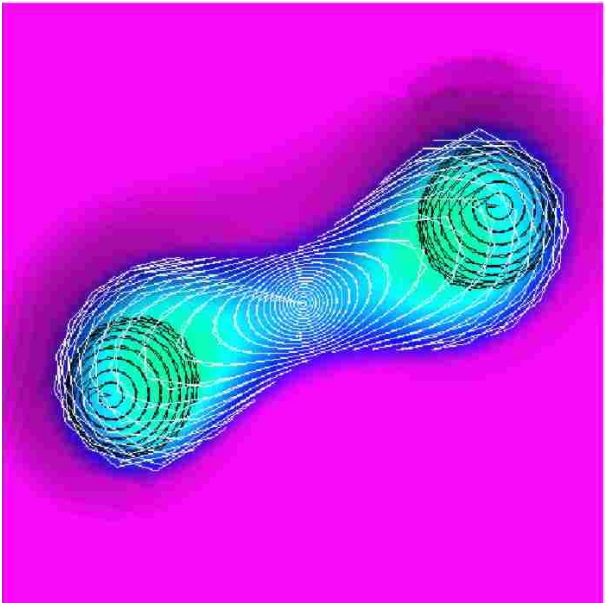}}
\vspace{0.5cm}
\caption{Level
set extraction for asymmetric binary black hole coalescence: II.}
\label{m:9}
\end{figure}

Figures  (\ref{m:8}) and  (\ref{m:9}) show two views of the
extracted level set $S = 0$ at $t=2M$. Note that this surface is highly
distorted and shows the event horizon just after merger. Also shown in
these figures are the apparent horizons for the two black holes.
In these figures each color of the color
map denotes a level set $\hat{\Gamma}$ of the eikonal. As such,
each color represents a null surface. From these figures it is
apparent that at this time there appears one innermost null
surface that completely contains both apparent horizons. Thus, the results
of the viscosity solutions suggest a merger time much closer to
$2M $ then the $8M$ found with the apparent horizon trackers.
[Note that in figures  (\ref{m:8}) and  (\ref{m:9}) the event horizon is 
shown as a white wire mesh, while the apparent horizons are shown
as black wire meshes. This is opposite to the color scheme used 
in figures \ref{m:1}, \ref{m:2}, \ref{m:5} - \ref{bh:8}, which was an 
independent study of the evolution as opposed to the study of the throat
geometry considered here.]

\section{Change of Topology}
\begin{figure} 
\epsfxsize=8cm
\centerline{\epsfbox{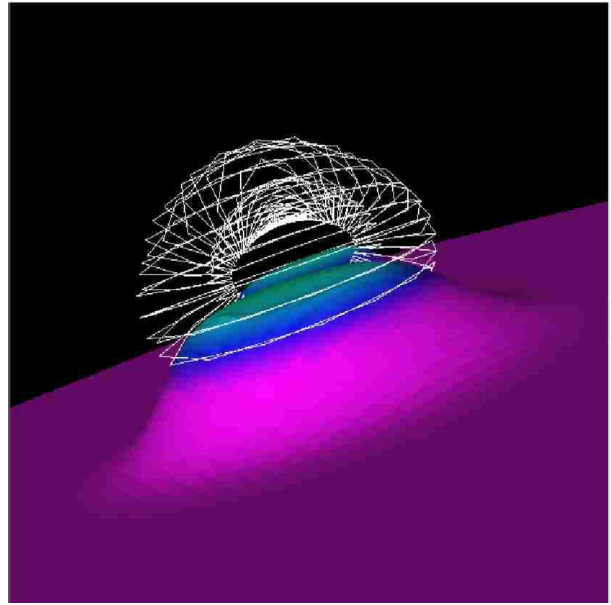}}
\vspace{0.5cm}
\caption{Throat
function for asymmetric binary black hole coalescence:
$t_{\uparrow} = 1.562 M $.} \label{m:10}
\end{figure}

As shown in figure (\ref{m:4}) the viscosity solutions of the
eikonal equation do continuously monitor a change in topology. In
the case of asymmetric binary black hole coalescence it is
conjectured that the level set $\Gamma$, which corresponds to a
section of the event horizon, must take a higher genus topology at
merger. To investigate this possibility, figure (\ref{m:10}) shows
the level set $\hat{\Gamma}$ viewed along the axis joining the
centers of the apparent horizons. In that figure it is apparent
that the throat function of the topological transition is
elliptical in geometry. Studies indicate that this elliptical
throat function persists for all null surfaces (i.e. those 
slightly inside or slightly outside our candidate event horizon) undergoing the
topological transition. Further, for all of our computed transitions of the
null surfaces, no higher genus topology is exhibited; instead,
the elliptical geometry of the throat persists to the transition.
These results suggest that if there is a non trivial topology in
the sections of the event horizon as a consequence of the
asymmetry of the merger, then that topology change is bounded to occur
when the minor axis of the ellipse is within one of our computational 
zones, or $h = M /4$.

\section{Area Analysis}
\begin{figure} 
\epsfxsize=8cm
\centerline{\epsfbox{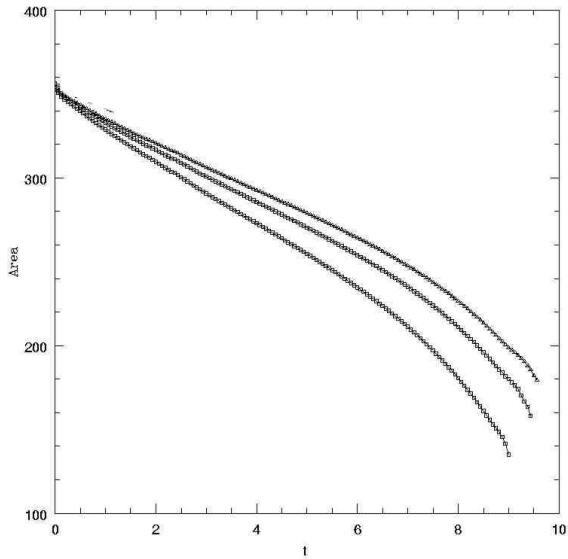}}
\vspace{0.5cm}
\caption{Area
versus time for asymmetric binary black hole coalescence.The horizontal 
scale is $t_{\downarrow}$, i.e. time measured into the past. 
The curves are (bottom to top) for 
$\epsilon = 2 h^{2}$, $\epsilon = h^{2}$, and the improved viscosity
solution.}
\label{n:1}
\end{figure}

\section{Black Hole Areas}

Figure  (\ref{n:1}) shows an area versus time plot for this
asymmetric collision. The curve with
the lowest area is the result of a viscosity solution with
$\epsilon = 2 h^2$. The curve with the second lowest area is the
result of a viscosity solution with $\epsilon = h^{2}$, while the
topmost curve is an improved viscosity solution
composed of the two higher viscosity solutions. 
According to these results it is immediately apparent that the
viscosity solutions of higher viscosity show a merger time that is
later in $t_{\uparrow}$ (and therefore prior in $t_{\downarrow}$)
then the merger time found with solutions constructed in  the
limit of vanishing viscosity. These results then indicate that the
error (or bias) in the merger time of the viscosity solutions is
directly related to the magnitude of the viscosity. More
precisely, for the continuum merger time of $t^{*}_{\uparrow}$ and
an approximate merger time of ${t^{*}_{\uparrow}}_{\epsilon}$,
constructed using a viscosity solution of viscosity parameter
$\epsilon$, the function $f\left(\epsilon\right) =
{t^{*}_{\uparrow}}_{\epsilon} - t^{*}_{\uparrow}$ is increasing in
$\epsilon$. (Here all
surface areas are calculated with  $m^{2}$ points where $m=100$.)

\begin{figure} 
\epsfxsize=8cm
\centerline{\epsfbox{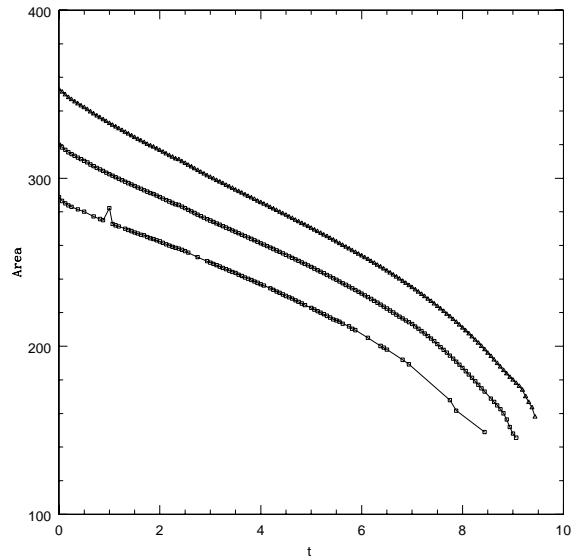}}
\vspace{0.5cm}
\caption{Areas
versus time for asymmetric binary black hole coalescence. The horizontal 
scale is $t_{\downarrow}$. }
\label{n:2}
\end{figure}

Figure (\ref{n:2}) shows several area versus time curves for
initial data of the form
\begin{equation}
u_{\delta}\left(\theta, \phi\right) = u_{c}\left(\theta,
\phi\right) - \delta
\end{equation}
where the data $u_{c}\left(\theta, \phi\right)$ is that obtained
using the method of section VII. From top to bottom, the curves show areas for $\delta
= 0, M/4, M/2$ and with a viscosity parameter of $\epsilon =
h^{2}$. Recall that studies of the apparent horizons found that
the true event horizon is contained in the domain $0 < \delta <
M/2$.  This survey over $\delta$ is conducted in search for the
convergence signature associated to event horizons. Due to the time
scale of this data, the time scale of the dynamics, and the
relaxation time scale of the event horizon tracking method, the
signature is not clearly apparent. However, this study of the area
curves does show convergence of the areas, which is expected for
null data approaching the horizon when propagated into the past.
The curve with $\delta = M/2$ shows behavior suggestive of an
event horizon since the $\delta \neq M/2$ curves all approach that of
$\delta =  M/2$. The $\delta =  M/2$ curve is therefore considered
the best candidate for the numerical event horizon of this study.
Note that the sections of this event horizon completely contain
the correct apparent horizons for all $t_{\downarrow} < 9M$.
Further, these $\delta = M/2$ data show a bifurcation time at $t_{\downarrow}
\approx 8.3 M$, which corresponds to a merger time of about
$t_{\uparrow} \approx 1.7 M$. This bifurcation time is detected by
an algorithm that searches for any points $\theta, \phi$ of the
surface such that $u\left(\theta, \phi\right) < h $. In the
circumstance that a point is found such that $u\left(\theta,
\phi\right) < h $, the bifurcation is expected to occur in a few
more $dt = M/16$ in the $t_{\downarrow}$ direction. A few time
levels prior to this bifurcation time in $t_{\downarrow}$ (i.e. just
after the bifurcation in $t$), the single merged horizon component 
has an area (the area of the level set) computed to be
$A = 148.9$. Based on analysis of area computations for exact solutions 
(figure(\ref{vserrora:1}), we anticipate an error in the horizon area of
several percent. We conservatively assign an $8\%$ error to the areas.
Just prior (in $t$) to merger the area of each black hole is
then $A = 74.5 \pm 6.0 M^2$. These individual areas correspond to a
Schwarzschild mass of about $M = 1.48 \pm 0.12$. This result is a
substantially larger mass for each hole then determined by
the apparent horizon finders at the time level $t_{\uparrow} = 0$.
Interestingly, studies have found that apparent horizons separated
by about $10M$ show an increase in their mass
due to the effects of binding energy\cite{bonning}. Individual masses of about
$M \approx 1.36$ are then only a slight departure from studies
that account for the binding energy of the holes. Further, at the
time of merger the holes have undergone $1.7 M$ of evolution,
during which the holes could accrete any surrounding gravitational
radiation present in the initial data. The presence of such
radiation would lead to larger masses then those found using
apparent horizon finders at $t_{\uparrow} = 0$. However, it is
important to note that due to the viscosity in the solution the
result $M \approx 1.36$ can only properly be considered as a lower
bound on the calculated masses. 
The most significant contribution to any error in this result
must stem from the relatively small time scale of this asymmetric
collision data and coupling of that time scale to the $e-$folding
time scale of this event horizon detection method.

\section{Conclusions}
In this work we have demonstrated a relatively 
simple yet robust and (most importantly) generic solution
to the problem of numerically tracking black hole event horizons.
An implementation of our method made use of an explicit second
order diffusion term to regulate the solution singularities
associated to caustics. As demonstrated by analysis of 
analytic sources, this term does introduce numerical 
error although we demonstrate our control over these effects and 
the resulting accuracy. But, the use of a second order
diffusion term is not required by our method per se; 
and a variety of other approaches can be employed. Examples of 
other methods for controlling breakdown of the numerical solution
include those classes of high resolution shock capturing
numerical schemes that are used extensively in computational 
fluid dynamics for hyperbolic problems similar to the eikonal equation.

The application of our new method  for event horizon tracking method
considered the asymmetric binary black hole coalescence problem,
including a detailed analysis of areas of the surfaces of sections,
the collision time, associated apparent horizons, and the topology of the 
horizon. Due
to the relatively short time scale of the collision data, our method
was unable to demonstrate the signature of the black hole
event horizon. We believe that this problem is due to the data 
itself and not due to our method. 
We anticipate much more accurate and convincing results as more
accurate computational simulations of black hole interactions 
become available. 



\begin{references}
\bibitem{he:}
  S. W. Hawking and G. F. R. Ellis,
  ``The Large Scale Structure of Spacetime",
  Cambridge University Press(1973).

\bibitem{opsny:}
  J. R. Oppenheimer  and H. Snyder,
  Phys. Rev.,
  {\bf 56},
  455
(1939).


\bibitem{carter:}
  B. Carter,,
  Phys. Rev. Lett.,
  {\bf 26},
  331
  (1971).
\bibitem{bch:}
  J. M. Bardeen, B.Carter, and S. W. Hawking,
  Commun. Math. Phys.,
  {\bf 31},
  161
(1973).


\bibitem{wald2:}
  R. M. Wald,
  ``Quantum Field Theory in Curved Spacetime and Black Hole Thermodynamics",
  University of Chicago Press
  (1994).


\bibitem{choptuik:}
  M. W. Choptuik,
  Phys. Rev. Lett.,
  {\bf 70},
  9
  (1993).






\bibitem{as1:}
A. Celotti, J. C. Miller, and D. W. Sciama,
Class. Quant. Grav.,
{\bf 16},
3	
(1999).

\bibitem{e1:}
K. Menou, E. Quataert, and R. Narayan,
Proc. $8^{th}$ Marcel Grossmann Meeting on General Relavity,
(1997).

\bibitem{ambh:}
Richstone \textit{et} \textit{al},
Nature,
{\bf 395},
A14
(1998).

\bibitem{eb:}
K. Belczynski, V. Kalogera, and T. Bulik,
Astrophys. J.,
{\bf 572},
407
(2001).

\bibitem{lehner:}
L. Lehner,
Class. Quantum. Grav.,
{\bf 18},
161
(2001).


\bibitem{bbhc1:}
L. P. Grischuk \textit{et} \textit{al},
Physics-Uspekhi,
{\bf 171},
3
(2001).


\bibitem{eh1:}
S. A. Hughes \textit{et} \textit{al},
Phys. Rev D.,
{\bf 49},
4004
(1994).

\bibitem{eh2:}
P. Anninos \textit{et} \textit{al},
Phys. Rev Lett.,
{\bf 74},
630
(1995).


\bibitem{eh3:}
R. A. Matzner \textit{et} \textit{al},
Science,
{\bf 270},
941
(1995).

\bibitem{eh4:}
J. Libson \textit{et} \textit{al},
Phys. Rev. D.,
{\bf 53},
4335
(1996).

\bibitem{eh5:}
J. Masso \textit{et} \textit{al},
Phys. Rev. D.,
{\bf 59},
(1999).

\bibitem{is1:}
A. Ashtekar \textit{et} \textit{al},
Phys. Rev. Lett.,
{\bf 85},
3564
(2000).

\bibitem{newman1:}
J. Ehlers and E. T. Newman,
J. Math. Phys.,
{\bf 41},
3344
(2000).

\bibitem{newman2:}
S. Frittelli and E. T. Newman,
J. Math. Phys.,
{\bf 40},
383
(1999).

\bibitem{newman3:}
E. T. Newman and A. Perez,
J. Math. Phys.,
{\bf 40},
1093
(1999).

\bibitem{icn:}
S. A. Teukolsky,
Phys. Rev. D,
{\bf 61},
087501
(2000).

\bibitem{wini2:}
S. Husa and J. Winicour,
Phys. Rev D.,
{\bf 60},
84019
(1999).

\bibitem{exc1:}
S. Brandt \textit{et} \textit{al},
Phys. Rev. Lett.,
{\bf 85},
5496
(2000).



\bibitem{deirdre:}
D. M. Shoemaker, M. F. Huq, and R. A. Matzner,
Phys. Rev. D,
{\bf 62},
124005
(2000).

\bibitem{bonning}E. Bonning, D. Neilsen, and Richard A. Matzner,
"Physics an dInitial Data for Black Hole Spacetimes", in preparation
(2003).

\end{references}
\end{document}